\documentclass[aps,pra,floatfix,superscriptaddress,showpacs,twocolumn]{revtex4}
\usepackage{amsmath}
\usepackage{graphicx}

\newcommand{\abs}[1]{\left|#1\right|}

\begin{document}
\title{Magnetometry with millimeter-scale anti-relaxation-coated alkali-metal vapor cells}
\author{M. V. Balabas}
\affiliation{S. I. Vavilov State Optical Institute, St. Petersburg,
199034 Russia}
\author{D. Budker}
\email{budker@berkeley.edu} \affiliation{Department of Physics,
University of California, Berkeley, CA 94720-7300}
\affiliation{Nuclear Science Division, Lawrence Berkeley National
Laboratory, Berkeley CA 94720}
\author{J. Kitching}\email{kitching@boulder.nist.gov}
\affiliation{National Institute of Standards and Technology, 325 S.
Broadway, Boulder, CO 80305-3322}
\author{P. D. D. Schwindt}
\affiliation{National Institute of Standards and Technology, 325 S.
Broadway, Boulder, CO 80305-3322}
\author{J. E. Stalnaker}
\affiliation{Department of Physics, University of California,
Berkeley, CA 94720-7300} \affiliation{National Institute of
Standards and Technology, 325 S. Broadway, Boulder, CO 80305-3322}

\date{\today}
\begin{abstract}
Dynamic nonlinear magneto-optical-rotation signals with frequency-
and amplitude-modulated laser light have been observed and
investigated with a spherical glass cell of 3-mm diameter containing
Cs metal with inner walls coated with paraffin. Intrinsic Zeeman
relaxation rates of $\gamma/(2\pi)\approx 20\ $Hz and lower have
been observed. Favorable prospects of using millimeter-scale coated
cells in portable magnetometers and secondary frequency references
are discussed.
\end{abstract}


\maketitle

\section{Introduction}

The state of the art in compact atomic frequency references is
represented by devices about 100~cm$^3$ in volume, dissipating
several watts of power \cite{Compact_Clocks}, and maintaining
relative frequency stability of about 10$^{-11}$ over one day.
Current atomic magnetometers can achieve high sensitivity (see, for
example, Refs. \cite{Ale2003_OP}, \cite{Bud2000Sens,Kom2003}) but
are typically large and nonportable or cumbersome to carry.
Recently, physics packages for highly compact clocks
\cite{Kna2004,Lut2004} and magnetometers \cite{Sch2004} have been
developed that are based on microelectromechanical systems (MEMS)
micromachining techniques and promise to allow even further
reduction of the size and power of these types of instruments.

Until now, all of these compact devices have been based on
alkali-vapor cells containing a buffer gas. However the use of
anti-relaxation coated cells in such devices may have a number of
benefits, including a reduced sensitivity to field gradients and a
significantly narrower resonance linewidth at very small size scales
(see, for example, Ref. \cite{Kit2002}).




With anti-relaxation coatings, polarized alkali atoms (Cs, Rb, K)
can experience multiple wall collisions (up to $10^4$) without
depolarization. Recent experimental investigations
\cite{Bud2005NIST,Guz2005} have explored Zeeman and hyperfine
relaxation in rubidium- and potassium-metal-vapor cells with
different paraffin coatings. The characteristic cell dimensions in
those works were 3 to 10$\ $cm.

Here we investigate nonlinear magneto-optical signals  with
millimeter-scale paraffin-coated cells with cesium. We verify the
scaling of the relaxation properties with cell size for
millimeter-scale cells and discuss the practical aspects of
magnetometric measurements with such cells.

We have found the 3-mm cells appropriate for small-scale
magnetometers (and, probably, clocks), and verified that the
intrinsic Zeeman-relaxation rate at a given alkali-vapor pressure
scales approximately as $1/D$, where $D$ is the characteristic
dimension of the cell, for cell sizes varying by 1.5 orders of
magnitude. We also compared nonlinear magneto-optical rotation
(NMOR) signals using two different types of laser modulation --
frequency modulation (FM NMOR \cite{Bud2002FM}) and amplitude
modulation -- AM NMOR. We find that the AM NMOR signals are
comparable to the FM NMOR signals in terms of width and signal
strength, suggesting that the method of choice may depend on the
ease of implementation in a particular experiment or device.
Finally, the NMOR signals observed in this work at relatively high
light power show evidence of conversion between alignment created in
atoms by the laser light and other polarization moments such as
orientation.

\section{The cell}
\begin{figure}
\includegraphics[width=3.4 in]{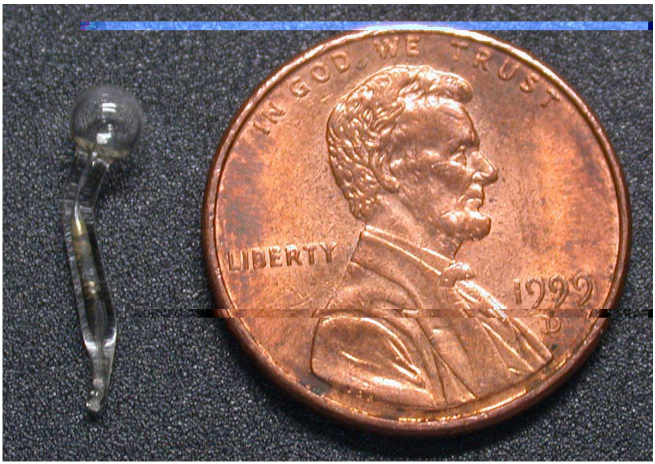}
\caption{Cs-vapor microcell, whose inner walls are coated with
paraffin, used for the present
measurements.}\label{Fig_Cs_microcell}
\end{figure}

The cell used for the measurements reported here is shown in Fig.\
\ref{Fig_Cs_microcell}. The general technology for cell preparation
and application of the paraffin coating is described in Ref.
\cite{AleLIAD}. The cell was made by glass-blowing techniques and is
of an approximately spherical shape, with an outer diameter of about
3$\ $mm and wall thickness of about 0.2 to 0.3$\ $mm. A cesium-metal
sample is placed in a stem (made at the same time as the cell) that
is connected to the cell via a thin capillary about 0.2$\ $mm in
diameter and about 2$\ $mm in length. The broader section of the
stem on the side of the capillary opposite to the cell contains Cs
metal and is about 6$\ $mm long.

\section{Apparatus, procedure, and results}
Most of the measurements reported here were performed by the FM NMOR
technique \cite{Bud2002FM}. Recent reviews of nonlinear
magneto-optics of resonant vapors are given in Refs.
\cite{Bud2002RMP,Ale2005}. The vapor cell at room temperature was
mounted within a cylindrical double-layered magnetic shield. We used
a distributed-feedback (DFB) laser producing up to 10$\ $mW of cw
light, tuned to the Cs D1 line (with wavelength in vacuum of
$\lambda=894.6\ $nm). The laser beam was attenuated, passed through
a Glan polarizer, and directed through holes in the magnetic shield
through the vapor cell. The light beam was apertured to about 2$\
$mm in diameter before entering the cell. The curvature of the cell
caused the transmitted light beam to diverge. This divergence was
compensated by installing a 2.5$\ $cm-diameter anti-reflection
coated lens with a focal length of 2.5$\ $cm at a distance of 0.15$\
$cm after the cell. The light after the lens exited the magnetic
shield and was analyzed with a Wollaston polarizer and a balanced
photoreceiver whose output was connected to a
digital-signal-processing lock-in amplifier. The laser frequency was
sinusoidally modulated by dithering the junction current, and the
synchronous signal due to optical rotation was detected.

From auxiliary absorption measurements, we estimated the Cs-vapor
density in the cell to be approximately $3\cdot 10^9\ $cm$^{-3}$.
This density is an order of magnitude lower than the saturated
density of Cs at room temperature, in our experience, a
not-too-uncommon occurrence in paraffin-coated cells (the density
can usually be brought closer to the saturated density by reheating
the stem; however, this was not needed here, as the signals were
sufficiently strong even with this low alkali-vapor density). A
solenoid was installed within the innermost magnetic shield and was
used to set or scan the magnetic field applied to the cell directed
collinearly with the light-propagation direction (the
Faraday-rotation geometry). In a typical measurement, we would tune
the central frequency of the laser to a particular value near the
resonance, set the modulation frequency and amplitude, and record
the output of the lock-in amplifier as a function of the magnetic
field.

\begin{figure}
\includegraphics[width=3.4 in]{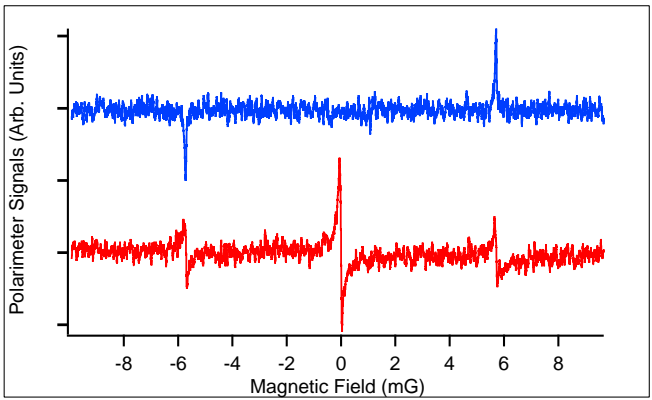}
\caption{An example of a low-light-power magnetic-field scan. Lower
trace is the in-phase, and the upper trace is the quadrature signal.
The modulation frequency was $\Omega_M=2 \pi\cdot 4\ $kHz. The depth
of the frequency modulation was $\Delta \nu = 750\ $MHz
peak-to-peak. The total scan time was 10$\ $s. The laser power was
0.12$\ \mu$W. The resonance widths for this scan are $\delta B
\approx 55\ \mu$G, which corresponds to $\gamma\approx 2\pi \cdot
19\ $Hz. The center frequency of the laser is tuned to the
low-frequency slope of the D1 $F=4\rightarrow F'=3$
resonance.}\label{Fig_LowPowerFMScan}
\end{figure}

An example of such an FM NMOR scan is shown in Fig.\
\ref{Fig_LowPowerFMScan}. For this scan, the light power transmitted
through the cell was $\approx 0.12\ \mu$W, and the modulation
frequency of the laser was set to $\Omega_M=2\pi\cdot 4\ $kHz, with
the peak-to-peak frequency-modulation depth of $\Delta \nu = 750\
$MHz. The upper trace on the figure shows the quadrature
(out-of-phase) output of the lock-in amplifier, while the lower
trace (which is vertically offset for clarity) represents the
in-phase output. As usual in FM NMOR experiments, there is a
dispersively shaped resonance in the in-phase component that appears
near zero magnetic field, as well as dispersively shaped resonances
that appear when $\Omega_M=2\, \Omega_L$, where $\Omega_L=\abs{g\mu
B}$ is the Larmor-precession frequency, $g$ is the Land\'{e} factor
of the ground-state hyperfine component ($g=\pm 1/4$ for the Cs
$F=4,3$ states, respectively), $\mu$ is the Bohr magneton, and $B$
is the magnetic field. Absorptively shaped quadrature resonances
also appear when $\Omega_M= 2\, \Omega_L$. The origin of all these
features is well understood qualitatively \cite{Bud2002FM}, and they
are reproduced in an analytical calculation with a model system
\cite{Mal2004}.

\begin{figure}
\includegraphics[width=3.4 in]{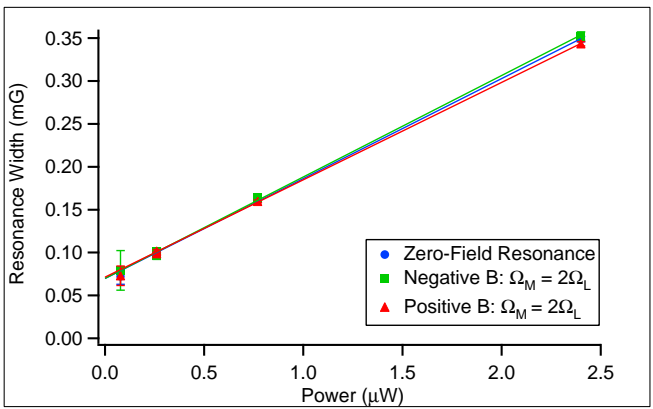}
\caption{Light-power dependence of the resonance widths. The
extrapolation towards zero light power for this data set yields an
intrinsic relaxation rate of $\gamma/(2\pi) \approx 24\ $Hz,
consistent for the zero- and nonzero-field resonances.}
\label{Fig_PowerDep}
\end{figure}

We use the scans such as the one shown in Fig.\
\ref{Fig_LowPowerFMScan} to determine the Zeeman-relaxation rate
$\gamma$. To do this, we fit the resonances to appropriately phased
Lorentzians and determine their width $\delta B$. The width $\delta
B$ corresponds to the separation between the minimum and the maximum
of the dispersively shaped features in the in-phase signals. We then
evaluate the relaxation rate from $\gamma = 2 \pi\, |g| \mu\, \delta
B$. The rate $\gamma$ has a contribution due to light-power
broadening. In order to find the intrinsic light-independent
relaxation rate, we take data at several low light-power levels, and
extrapolate to zero light power (Fig.\ \ref{Fig_PowerDep}). The data
shown in Fig.\ \ref{Fig_PowerDep} were taken over a period of less
than an hour on one day. The intrinsic relaxation rate obtained from
these data is $\gamma/(2\pi)=24.4(5)\ $Hz (corresponding to the
resonance width as a function of the magnetic field of 0.07~mG). On
a different day (on which the scan of Fig. \ref{Fig_LowPowerFMScan}
was taken), we observed smaller intrinsic relaxation rates, down to
10 to 15$\ $Hz. We attribute this variation to the change in the
equilibrium vapor density in the cell. Indeed, previous work
\cite{Bud2005NIST,Gra2005} has produced evidence that wall
relaxation of polarized atoms in paraffin-coated cells is strongly
dependent on the vapor density in the cell.

\begin{figure}
\includegraphics[width=3.4 in]{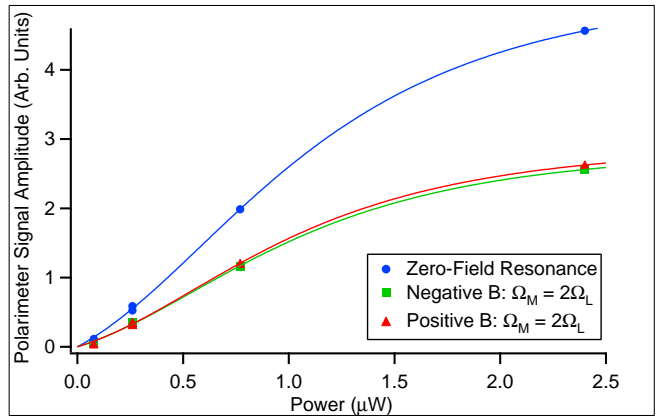}
\caption{Light-power dependence of the signal amplitudes. The data
points match those in Fig. \ref{Fig_PowerDep}. The data are fit to
saturation curves with expected quadratic low-power
asymptotics.}\label{Fig_AmpVsPowerDay1}
\end{figure}

Figure \ref{Fig_AmpVsPowerDay1} shows the dependence of the signal
size vs. light power. The data points correspond to those in Fig.\
\ref{Fig_PowerDep}. The power dependence of the signal in the
low-power limit is expected to be quadratic because the signal is
proportional to the light power times the angle of the nonlinear
light-polarization rotation, which is in turn proportional to the
power. As seen in Fig.\ \ref{Fig_AmpVsPowerDay1}, significant
deviations from this asymptotic low-power behavior occur at
essentially all light powers where data were taken, and certainly at
$\sim 0.5\ \mu$W, where the dependence appears linear. Similar
saturation behavior was seen in earlier work on FM NMOR with Rb
\cite{Yas2003Select}.

Further details of the investigation of FM NMOR in this system are
given in the Appendices.

\section{Specifics of magneto-optics with small coated cells; estimate of sensitivity}
The results described above show that strong nonlinear
magneto-optical-rotation signals appropriate for use in
magnetometers can readily be obtained in millimeter-scale
paraffin-coated vapor cells (Fig.\ \ref{Fig_CalibratedHighPowerScan}
shows a spectral scan where the signals have been normalized, so the
first-harmonic amplitudes of the optical rotation are presented). We
point out that Cs has an advantage for use in magnetometers based on
atomic alignment (and higher polarization moments) because its
hyperfine structure is fully resolved (due to large upper-state
hyperfine splitting of $1168\ $MHz in the $6\,^2P_{1/2}$ state and
smaller Doppler width compared to lighter alkali atoms). As will be
discussed in more detail in Ref.\ \cite{Guz2005}, there is
considerable suppression of nonlinear magneto-optical rotation when
upper-state hyperfine structure is not fully resolved (a
particularly strong suppression, by two orders of magnitude compared
to cesium, in the case of potassium).

A concern in using small cells in magneto-optical work is the effect
of optical imperfection of glass-blown cells and the curvature of
the cell's surfaces. We found that, although we needed to use a lens
to correct for the cell's curvature, this does not present a serious
problem. We also found that, presumably due to stress-induced
birefringence in the glass, linearly polarized light acquired
significant ellipticity (up to $\sim 0.1\ $rad) upon propagation
through the cell. Ellipticity could lead to offsets in the
polarimeter signals due to self-rotation \cite{Roc2001SR}. We have
also observed that the ellipticity causes phase shifts for nonzero
B-field resonances and an asymmetry between positive and negative
B-field resonances at relatively high light powers (Fig.\
\ref{Fig_CalibratedHighPowerScan}). We were able to affect and
largely compensate these distortions by placing a quarter-wave plate
before the cell and adjusting it to compensate the ellipticity. This
problem could be eliminated completely through the use of planar,
microfabricated cells as described in \cite{Lie2004}. These cells
have windows made from thin, flat glass wafers and are expected to
have significantly less birefringence than the cell used in this
experiment.

While no systematic optimization or characterization of the
magnetometer performance was attempted in the present work, we have
measured that the setup in the present incarnation is a magnetometer
with a sensitivity to low frequency ($<50\ $Hz) variations in the
magnetic field of $\sim 4\ $pT/$\sqrt{\rm{Hz}}$. For this
measurement, the response of the system was determined by stepping
the frequency of the laser-frequency modulation by 1 or $2\ $Hz and
measuring the voltage change at the in-phase output of the lock-in
amplifier. This gives the slope as a function of frequency, which is
then converted to a slope as a function of magnetic field. Noise was
evaluated by connecting the output of the lock-in amplifier to a
spectrum analyzer. The magnetic sensitivity was measured with laser
power of $310\ $nW, peak-to-peak frequency-modulation amplitude of
$992\ $MHz, and the laser center frequency tuned to maximize the
slope, $\approx -500\ $MHz from the $F=4\rightarrow F'=3$
transition.

\section{Nonlinear magneto-optical rotation with amplitude-modulated light (AM NMOR)}
We have also briefly explored an alternative technique to FM NMOR,
where instead of frequency modulating the laser synchronously with
the Larmor precession, laser-light amplitude is modulated instead
(AM NMOR). In fact, this technique of synchronous optical pumping is
similar to that described in the pioneering work of Bell and Bloom
\cite{Bel61a}. Important differences with the early work include
that we pump transverse atomic alignment rather than a combination
of longitudinal alignment and orientation, and that we synchronously
detect optical rotation rather than transmission (see review article
\cite{Ale2005} for a discussion of various dynamic nonlinear
magneto-optical processes and references to related work).

\begin{figure}
\includegraphics[width=3.4 in]{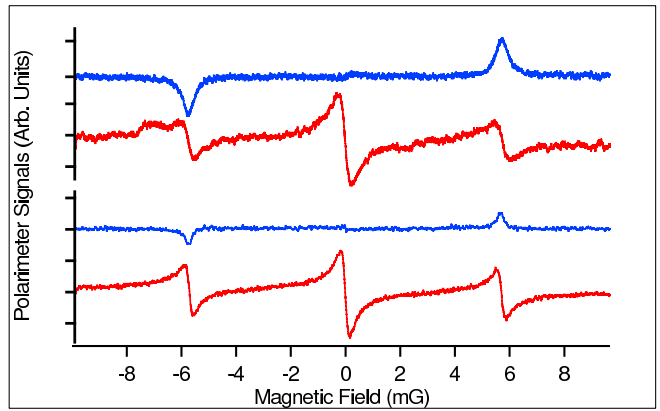}
\caption{A comparison of AM (upper plot) and FM NMOR (lower plot)
signals taken in similar conditions. These data were taken at light
power (maximum light power in the AM case) of $\approx 1.8\ \mu$W
with modulation frequency of 4$\ $kHz in both cases. Small
dispersive features at zero magnetic field are due to imperfect
phasing of the lock-in amplifier.}\label{Fig_AMvsFM_NMOR}
\end{figure}

To achieve amplitude modulation of the laser power, the light beam
was passed through a commercial acousto-optical modulator-deflector
(AOM) operated at the acoustic frequency of $80\ $MHz. The drive of
the AOM could be controlled by applying a TTL signal from a function
generator. To compare AM and FM NMOR signals, we recorded both types
of signals at the same modulation frequency, and otherwise similar
experimental conditions (Fig.\ \ref{Fig_AMvsFM_NMOR}). For the AM
data shown, we used square-wave (on/off) modulation of the laser
power with a $50\ \%$ duty cycle.
Comparison of the AM and FM NMOR data shown in Fig.\
\ref{Fig_AMvsFM_NMOR} shows that both modulation methods produce
similar signals.

An interesting result of the comparison is that the quadrature
signals are more significantly suppressed with respect to the
in-phase signals in the case of FM NMOR. We hypothesize that this is
due to less efficient conversion of the induced atomic alignment
into other polarization multipoles due to a factor-of-two lower
average light power to which the atoms are exposed in the AM case.
In general, the AM and FM NMOR spectra are different. Consequently,
the relative merits of the two methods for magnetometry will have to
be assessed by a careful investigation and separate optimization of
various parameters involved in the measurement (light power, tuning
of the central light frequency, AM duty cycle, etc.). This work is
in progress elsewhere \cite{Gaw2005AMOR}.

As seen in Fig.\ \ref{Fig_AMvsFM_NMOR}, the AM NMOR signals are
noisier than their FM NMOR counterparts, particularly in the case of
the in-phase signal. The origin of this is not presently understood.
We suspect that this is due to elevated sensitivity to
laser-frequency noise, a point that requires further study.

\section{Conclusion}
We have conducted measurements of nonlinear magneto-optical-rotation
signals with frequency- and amplitude-modulated light (FM and AM
NMOR) with millimeter-scale paraffin-coated Cs cells. The results
indicate that such cells are promising for the development of
small-scale magnetometers. While no systematic optimization or
characterization of the magnetometer performance was attempted in
the present work, the measured sensitivity of $\sim 4\
$pT/$\sqrt{\rm{Hz}}$ suggests that sub-picotesla magnetometry might
be possible with highly miniaturized devices.
Since atomic magnetometers and clocks are essentially similar
devices, we also expect that coated cells with volumes comparable to
the one studied in this work could find application in secondary
frequency standards.


\section*{Acknowledgements}

The authors are grateful to E. B. Alexandrov, M. Auzinsh, W. Gawlik,
J. S. Guzman, L, Hollberg, D. F. Kimball, S. Pustelny, S. M.
Rochester, J. Zachorowski, and V. V. Yashchuk for helpful
discussions. This work was supported by the ONR-MURI grant No.
FD-N00014-05-1-0406, by the National Science Foundation, by the
Director, Office of Science, Office of Basic Energy Sciences,
Nuclear Science Divisions, of the U.S. Department of Energy under
contract DE-AC03-76SF00098,  by a CalSpace Minigrant, and by the
Microsystems Technology Office of the Defence Advanced Research
Projects Agency (DARPA). This work is a partial contribution of
NIST, an agency of the United States government, and is not subject
to copyright.

\section*{Appendix I. FM NMOR Spectra}
\begin{figure}
\includegraphics[width=3.4 in]{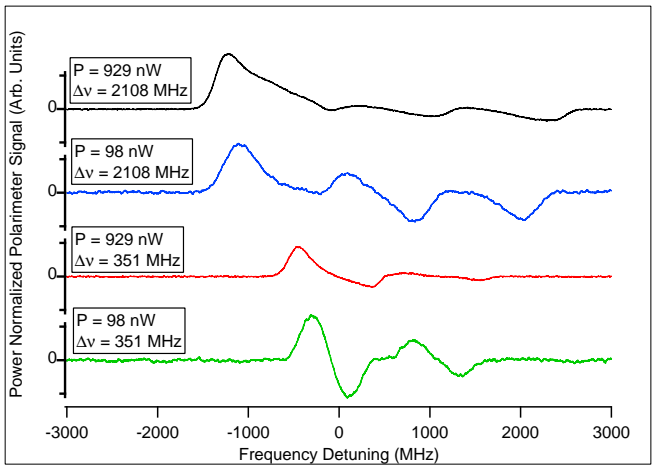}
\caption{Examples of FM NMOR spectra recorded at two different light
powers and for two peak-to-peak frequency-modulation amplitudes. The
modulation frequency was $2\ $kHz. The shown spectra are for the
in-phase components of the negative-B resonance for the $F=4
\rightarrow F'=3,4$ transitions. The energy separation between the
$F'=3,4$ levels corresponds to a frequency of $1168\ $MHz.
Vertical scales are the same for all
plots.}\label{Fig_ModPowerSpectra}
\end{figure}

We have investigated the spectra of FM NMOR signals for different
modulation amplitudes and light powers (Fig.\
\ref{Fig_ModPowerSpectra}). In order to remove the contributions to
the spectra that do not have sharp resonant character with respect
to magnetic field, we have subtracted spectra recorded at the
minimum and maximum of the negative-B feature. The nonresonant
contributions were typically at a level $\sim 30-50\ \%$ of the
resonant signal. These contributions are related to such effects as
etaloning on the optical elements, residual misbalance of the
polarimeter, etc.
The plots in Fig.\ \ref{Fig_ModPowerSpectra} indicate some
variations in the shape of the spectrum from low to high powers, and
a more dramatic variation with the modulation amplitude. While the
spectrum for low frequency-modulation amplitudes is composed of
dispersively shaped features corresponding to the different
hyperfine components of the transition, the
high-modulation-amplitude spectra are mostly absorptively shaped.
Qualitatively, this is because at low modulation amplitudes, the
spectrum is similar to the derivative of the non-modulated NMOR (see
Ref. \cite{Bud2002FM}), while at high modulation amplitudes, the
peaks occur when the central frequency of the laser is detuned in
such a way that the the laser frequency is in resonance with the
atoms at either the maximum (positive peaks) or the minimum
(negative peaks) of its frequency excursion. This, in a sense, is
equivalent to low-duty cycle (short-pulse) interaction of the laser
light with the atoms synchronous with the Larmor precession. The
peaks corresponding to a given hyperfine component are of opposite
signs for positive and negative detuning because optical rotation is
maximal at opposite phases of the laser-frequency modulation.

\begin{figure}
\includegraphics[width=3.4 in]{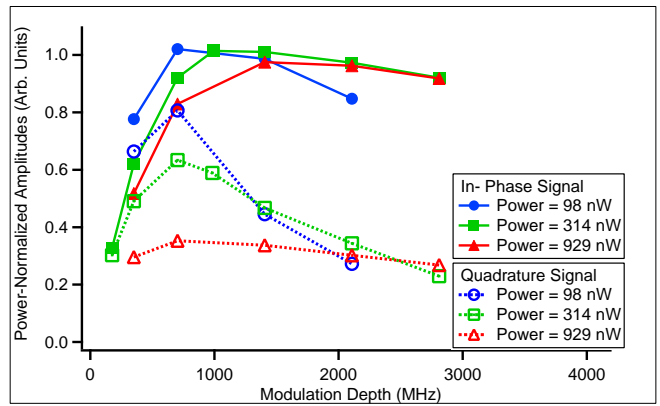}
\caption{Relative signal-peak amplitudes vs. peak-to-peak modulation
amplitude (modulation depth $\Delta \nu$) for different light powers
obtained from the in-phase spectra (as in Fig.
\ref{Fig_ModPowerSpectra}) and quadrature spectra. The experimental
points are connected only to guide the
eye.}\label{Fig_PetersAmpl_vs_ModAmpl}
\end{figure}

For practical applications of FM NMOR, it is important to optimize
various parameters of the system, including light power, central
laser-frequency detuning, modulation amplitude, etc. The figure of
merit for the magnetometer performance depends on the signal size
and resonance linewidth. While a full optimization was not attempted
here, Fig.\ \ref{Fig_PetersAmpl_vs_ModAmpl} shows the dependence of
the maximum signal on the modulation amplitude at different light
powers. To obtain this dependence, the modulation frequency was
fixed, and the central laser frequency was scanned to obtained
spectra similar to those shown in Fig.\ \ref{Fig_ModPowerSpectra}.
The plots in Fig. \ref{Fig_PetersAmpl_vs_ModAmpl} show the
amplitudes of the largest (low-frequency; see Fig.\
\ref{Fig_ModPowerSpectra}) peaks in the spectra normalized by light
power.
\begin{figure}
\includegraphics[width=3.4 in]{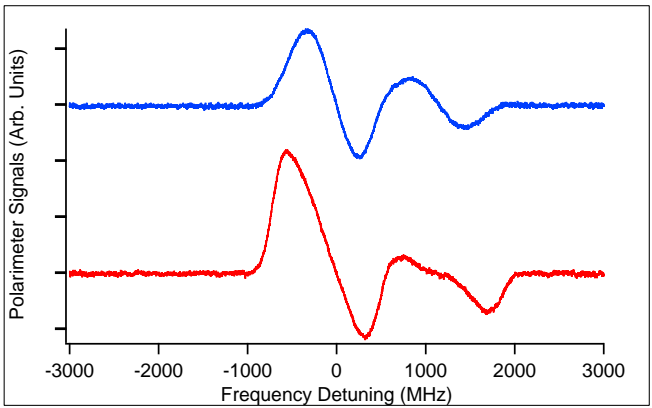}
\caption{An illustration of the difference in the spectra for the
quadrature (upper trace) and in-phase (lower trace) signals. Data
were taken at light power of $314\ $nW and peak-to-peak
frequency-modulation amplitude of $993\ $MHz and modulation
frequency of $2\ $kHz. The shown spectra are for the negative-B
resonance for the $F=4 \rightarrow F'=3,4$
transitions.}\label{Fig_IPQuadSpecM993P314}
\end{figure}

Figures \ref{Fig_ModPowerSpectra} and \ref{Fig_IPQuadSpecM993P314}
show the parts of the FM NMOR spectra corresponding to the
$F=4\rightarrow F'$ transition group. We have also recorded spectra
for the $F=3\rightarrow F'$ transition group. The signals for the
$3\rightarrow 3$ transition are several times smaller than those for
$F=4\rightarrow F'$, while the signals for the $3\rightarrow 4$
transition can hardly be seen at all at our sensitivity. A similar
distribution of the FM-NMOR-signal strength over the hyperfine
components of the D1 transition was observed in Rb \cite{Bud2002FM}.

\begin{figure}
\includegraphics[width=3.4 in]{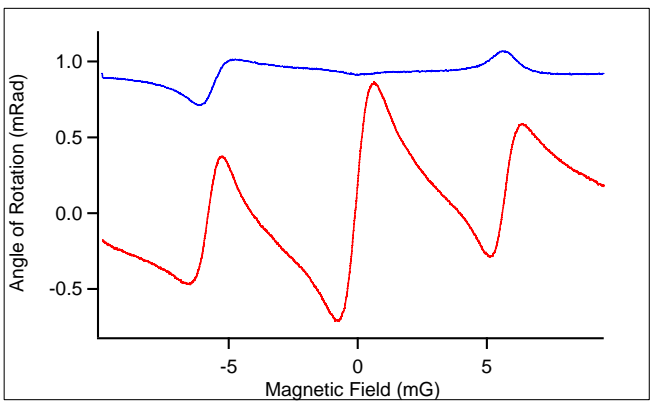}
\caption{A calibrated high-light-power (11.7$\ \mu$W) magnetic-field
scan. The FM NMOR signals are normalized, so the signals are given
in terms of the amplitude of the first harmonic of the optical
rotation. Upper trace -- quadrature signal; lower trace -- in-phase
signal. The laser frequency was modulated at $4\ $kHz with a
peak-to-peak modulation amplitude of 750$\ $MHz. The amplitude of
the quadrature signal is significantly smaller than that of the
in-phase signal. The positive- and negative-$B$ resonance signals
show considerable asymmetry (see text).
}\label{Fig_CalibratedHighPowerScan}
\end{figure}
\section*{Appendix II. Evidence for polarization-moment conversion}
At low powers, the quadrature and in-phase signals have similar
spectra; however, at higher powers, significant deviations are
observed (Fig.\ \ref{Fig_IPQuadSpecM993P314}). The differences
between the in-phase- and quadrature-signal spectra may be due to
the conversion of laser-induced atomic alignment into orientation
and other, higher, polarization moments (see reviews
\cite{Bud2002RMP,Ale2005} and references therein). Such conversion
occurs due to the combined action of the magnetic field and the ac
electric field of the light on the polarized atoms.

As a result of such polarization-moment conversion, components of
polarization are created that do not undergo Larmor precession (for
example, orientation directed along the magnetic field). Such
polarization components would not contribute to the synchronous
signals we detect if unmodulated probe light were used for
detection. (They would contribute to time-independent optical
rotation of the probe.) In the present case, however, the pump and
probe light are the same and are both modulated. Thus, this
``static" rotation also acquires a time dependence that is in phase
with the laser-frequency modulation. Since the spectra of optical
rotation for different polarization multipoles are, in general,
different, it is not surprising that these differences also manifest
themselves in the differences between the in-phase and quadrature
spectra.

Evidence for conversion of laser-induced alignment into other
polarization multipoles also comes from the fact that, while the
amplitudes of the in-phase and quadrature signals for the nonzero
B-field resonances are the same at low light powers, the relative
size of the quadrature signals decreases greatly at higher light
powers. For example, the quadrature-signal amplitudes are about a
factor of five smaller than the in-phase ones for the light power of
$\approx12\ \mu$W (see Fig.\ \ref{Fig_CalibratedHighPowerScan}).
This trend is also seen in the comparison of the quadrature and
in-phase amplitudes shown in Fig.\ \ref{Fig_PetersAmpl_vs_ModAmpl}.

\bibliography{NMObibl}

\end{document}